\documentclass[aps, amsmath, amssymb, 10pt, pra, twocolumn, longbibliography, superscriptaddress]{revtex4-1}

\usepackage{algcompatible}
\usepackage{amsmath}
\usepackage{amsthm}
\usepackage{bm}
\usepackage{braket}
\usepackage{subfigure}
\usepackage[dvipdfmx]{graphicx}
\usepackage{float}
\usepackage{url}
\usepackage{xcolor}
\usepackage{multirow}
\usepackage[colorlinks=true,urlcolor=blue,citecolor=blue,linkcolor=blue]{hyperref}
\usepackage{cleveref}
\usepackage{here}

\usepackage[noend]{algpseudocode}

\crefname{figure}{Fig.}{Figs.}
\crefname{equation}{Eq.}{Eqså.}
\crefname{section}{Sec.}{Sec.}
\Crefname{figure}{Figure}{Figures}
\Crefname{equation}{Equation}{Equations}
\Crefname{section}{Section}{Sections}

\DeclareMathOperator*{\argmin}{arg\,min}

\begin{document}

\preprint{APS/123-QED}

\title{Subspace variational quantum simulator}

\author{Kentaro Heya}
\email{kheya@qc.rcast.u-tokyo.ac.jp}
\affiliation{Research Center for Advanced Science and Technology (RCAST), The University of Tokyo, Meguro-ku, Tokyo 153-8904, Japan}

\author{Ken M. Nakanishi}
\affiliation{Institute for Physics of Intelligence, The University of Tokyo, Tokyo 113-0033, Japan}

\author{Kosuke Mitarai}
\affiliation{Graduate School of Engineering Science, Osaka University, 1-3 Machikaneyama, Toyonaka, Osaka 560-8531, Japan}
\affiliation{Center for Quantum Information and Quantum Biology, Osaka University, Japan}
\affiliation{JST, PRESTO, 4-1-8 Honcho, Kawaguchi, Saitama 332-0012, Japan}

\author{Zhiguang Yan}
\affiliation{RIKEN Center for Quantum Computing (RQC), Wako, Saitama 351-0198, Japan}

\author{Kun Zuo}
\affiliation{RIKEN Center for Quantum Computing (RQC), Wako, Saitama 351-0198, Japan}

\author{Yasunari Suzuki}
\affiliation{NTT Computer and Data Science Laboratories, NTT Corporation, Musashino 180-8585, Japan}
\affiliation{JST, PRESTO, 4-1-8 Honcho, Kawaguchi, Saitama 332-0012, Japan}

\author{Takanori Sugiyama}
\affiliation{Research Center for Advanced Science and Technology (RCAST), The University of Tokyo, Meguro-ku, Tokyo 153-8904, Japan}

\author{Shuhei Tamate}
\affiliation{RIKEN Center for Quantum Computing (RQC), Wako, Saitama 351-0198, Japan}

\author{Yutaka Tabuchi}
\affiliation{RIKEN Center for Quantum Computing (RQC), Wako, Saitama 351-0198, Japan}

\author{Keisuke Fujii}
\affiliation{Graduate School of Engineering Science, Osaka University, 1-3 Machikaneyama, Toyonaka, Osaka 560-8531, Japan}
\affiliation{Center for Quantum Information and Quantum Biology, Osaka University, Japan}
\affiliation{RIKEN Center for Quantum Computing (RQC), Wako, Saitama 351-0198, Japan}

\author{Yasunobu Nakamura}
\affiliation{Research Center for Advanced Science and Technology (RCAST), The University of Tokyo, Meguro-ku, Tokyo 153-8904, Japan.}
\affiliation{RIKEN Center for Quantum Computing (RQC), Wako, Saitama 351-0198, Japan}

\date{\today}

\begin{abstract}
Quantum simulation is one of the key applications of quantum computing, which accelerates research and development in the fields such as chemistry and material science.
The recent development of noisy intermediate-scale quantum~(NISQ) devices urges the exploration of applications without the necessity of quantum error correction.
In this paper, we propose an efficient method to simulate quantum dynamics driven by a static Hamiltonian on NISQ devices, named subspace variational quantum simulator~(SVQS).
SVQS employs the subspace-search variational quantum eigensolver~(SSVQE)~\cite{Nakanishi2018} to find a low-lying eigensubspace and extends it to simulate dynamics within the subspace with lower overhead compared to the existing schemes.
We experimentally simulate the time-evolution operator in a low-lying eigensubspace of a hydrogen molecule.
We also define the subspace process fidelity as a measure between two quantum processes in a subspace.
The subspace time evolution mimicked by SVQS shows the subspace process fidelity of $0.88$--$0.98$.
\end{abstract}

\pacs{Valid PACS appear here}
\maketitle

\section{Introduction}\label{sec:intro}
The dynamical properties of quantum systems are of great scientific interest and practically important for applications, and hence researchers have developed classical simulation algorithms such as the time-dependent density functional theory~\cite{Runge1984}.
A controllable quantum system must be advantageous in simulating such dynamics over a classical computer~\cite{Feynman1982}.
Simulations of quantum dynamics have been intensively studied as one of the most promising applications of quantum computers~\cite{RevModPhys.86.153}.

Methods for simulating quantum dynamics on a quantum computer were first proposed based on Trotterization~\cite{10.2307/2899535, Wiebe_2011, Poulin_2011} and have been extended by incorporating linear combinations of unitaries~\cite{2012Childs, PRXQuantum.2.010333}, truncated Dyson series~\cite{Berry_2015}, quantum random walk~\cite{Berry_2020, Campbell_2019}, and quantum signal processing~\cite{Low_2017}.
While these methods are based on rigorous algorithms with error guarantees, they run only on a fault-tolerant quantum computer and are not tolerant to device-specific errors without error correction.
The required circuits are also too deep to demonstrate their performance on the present NISQ devices.
While this problem is expected to be overcome in future by quantum error correction~\cite{bravyi1998quantum,fowler2012surface, hertzberg2020laserannealing}, assessment of the computational power of NISQ devices is also of great interest~\cite{Boixo2018, Bouland2018, Chen2018}.

A suitable approach for NISQ devices is quantum--classical hybrid algorithms that utilize a variational method and shallow parameterized quantum circuits~\cite{cerezo2021variational}.
These algorithms, such as variational quantum eigensolvers~\cite{Peruzzo2013, McClean2016, Bauer2016, Kandala2017}, generally do not guarantee the accuracy of solutions but can be feasibly implemented on NISQ devices.
Simulations of quantum dynamics in the framework of quantum--classical hybrid algorithms originated in variational quantum simulation~(VQS)~\cite{Li2016c, Yuan_2019, PhysRevLett.125.010501, McArdle_2019}.
VQS approximates an arbitrary time evolution within the representational capability of a parameterized quantum circuit by updating the variational parameters of the circuit at each time step according to the time-dependent variational principle.
While VQS applies to the simulation of time-dependent Hamiltonian dynamics, it requires experiments using ancilla qubits and controlled operations to update the variational parameters at each time step.
An easier way to solve the time-independent Hamiltonian dynamics without such an iterative procedure is to diagonalize the given Hamiltonian.
From the basis transformation matrix and eigenvalues obtained by diagonalization, the time-evolution operator can be easily reproduced.
However, diagonalizing a high-dimensional Hamiltonian generally requires the execution of a too-deep quantum circuit for NISQ devices.

Several hybrid quantum--classical algorithms have been proposed to solve time-independent Hamiltonian dynamics on NISQ devices~\cite{C_rstoiu_2020, otten2019noiseresilient, commeau2020variational, PRXQuantum.2.010333, gibbs2021longtime}.
A representative method is the so-called variational fast-forwarding~(VFF)~\cite{C_rstoiu_2020, gibbs2021longtime}.
In VFF, one first trains a parameterized quantum circuit to mimic the short time-evolution operator and then extrapolates the time evolution by updating the variational parameters, without calling for additional quantum experiments.
VFF fully approximates the time-evolution operator with a parameterized quantum circuit, which is as difficult as quantum optimal control with a large numbers of qubits.

In this paper, we propose a quantum--classical hybrid algorithm for efficiently simulating time-independent Hamiltonian dynamics and experimentally demonstrate it on superconducting qubits.
The method, named subspace variational quantum simulator~(SVQS), avoids the difficulties of Hamiltonian diagonalization described above by restricting itself to partial diagonalization.
First, we use subspace-search variational quantum eigensolver~(SSVQE)~\cite{Nakanishi2018}, to partially diagonalize only a low-lying eigensubspace of the Hamiltonian $\mathcal{H}$.
In SSVQE, we find a unitary $U(\bm{\theta}^*)$ that maps $k$ orthogonal input states $\{\ket{\varphi_i}\}_{i=1}^k$ chosen from the computational basis to the excited states up to the $k$th $\{\ket{E_i}\}_{i=1}^k$, each having the corresponding eigenenergy $E_i$.
It is expected that the representation capability of the parameterized quantum circuit required for a partial diagonalization of the Hamiltonian is much lower than that for the full diagonalization, and can be realized in a shallower circuit.
In SVQS, the inverse of the unitary, $U^\dagger(\bm{\theta}^*)$, maps each $\ket{E_i}$ to a computational basis $\ket{\varphi_i}$, where we can easily apply phase factors $e^{-i E_i t}$.
Unlike VFF, SVQS mimics the time-evolution operator only for states in a specific eigensubspace.
On the other hand, it is expected that the cost required for training parameterized quantum circuits is lower than VFF.
SVQS and VFF have a trade-off between the size of the simulatable eigensubspace and the easiness of the implementation, and thus are complementary with each other.

The rest of this paper is organized as follows.
In \cref{sec:method}, we describe the method of SVQS in detail.
In \cref{sec:experiment}, we demonstrate SSVQE and SVQS on a system with two superconducting transmon qubits to simulate quantum dynamics of a hydrogen molecule.
To characterize the subspace time evolution mimicked by SVQS, we define the subspace process fidelity as a measure between two quantum processes in a subspace.
Even with the limited device performance, the subspace time evolution in the low-lying eigensubspace of a hydrogen molecule mimicked by SVQS shows the subspace process fidelity of $0.88$--$0.98$.
SVQS is effective for experiments on NISQ devices because of the modest requirement for experimental devices.

\section{Methods}\label{sec:method}
In this section, we first introduce the subspace-search variational quantum eigensolver~(SSVQE)~\cite{Nakanishi2018}, which is the key ingredient for our proposal.
We then describe our proposal, the subspace variational quantum simulator~(SVQS).
In the following subsections, $\mathcal{H}$ denotes a $n$-qubit Hamiltonian transformed from the given Hamiltonian as follows:
\begin{align}
\mathcal{H}=\sum_{j=1}^{2^n} E_j \ket{E_j}\bra{E_j},
\end{align}
where $E_j$ and $\ket{E_j}$ are the $j$th eigenenergy and eigenstate of $\mathcal{H}$, respectively.
In applications for quantum chemistry, Jordan-Wigner~\cite{JORDAN} or Bravyi-Kitaev~\cite{BRAVYI2002210} transformation can be utilized to map a fermionic Hamiltonian to an $n$-qubit Hamiltonian.

\subsection{Subspace-search variational quantum eigensolver}
SSVQE is an algorithm for finding the $k$th or up to the $k$th excited states of Hamiltonian $\mathcal{H}$~\cite{Nakanishi2018}.
To find excited states up to the $k$th, SSVQE takes $k$ orthogonal states as inputs of a parametrized quantum circuit and minimizes the weighted sum of the expected energies of the output states.
The output states become automatically an orthogonal set by the conservation of orthogonality under the unitary transformation.
Therefore, we find all the excited states up to the $k$th via a single optimization procedure.
The procedure of SSVQE is summarized as follows:
\begin{enumerate}
    \item Construct a parameterized quantum circuit $U(\bm{\theta})$ and prepare $k$ initial states $\left\{\ket{\varphi_j}\right\}_{j=0}^k\ (k\leq 2^n)$, which are orthogonal to each other ($\braket{\varphi_i|\varphi_j}=\delta_{i,j}$).
    \item Minimize $\mathcal{L}_{\bm{\omega}}(\bm{\theta}) = \sum_{j=0}^k \omega_j \braket{\varphi_j|U^\dagger(\bm{\theta})\mathcal{H}U(\bm{\theta})|\varphi_j}$, where the weight vector $\bm{\omega}$ is chosen such that $\omega_i > \omega_j$ when $i < j$.
\end{enumerate}
Successfully optimizing $\bm{\theta}$ of an appropriate parameterized quantum circuit $U(\bm{\theta})$ by the procedure above, each output state $\ket{\psi_j(\bm{\theta})}\equiv U(\bm{\theta})\ket{\varphi_j}\ (j=0,1,\cdots,k)$ converges to the following state
\begin{align}
\ket{\psi_j (\bm{\theta}^*)}=e^{i\delta_j}\ket{E_j},
\end{align}
where $\delta_j$ is an unknown global phase factor, and $\bm{\theta}^*$ denotes the parameters that minimizes $\mathcal{L}_{\bm{\omega}}(\bm{\theta})$.
Therefore, the obtained circuit $U(\bm{\theta}^*)$ corresponds to a map between the computational subspace $\mathcal{S}_{\mathrm{com}}$ spanned by the orthogonal initial states $\{\ket{\varphi_j}\}_{j=1}^{k}$ and the eigensubspace $\mathcal{S}_{\mathrm{\parallel}}$ spanned by the excited states $\{\ket{E_j}\}_{j=1}^{k}$.

\subsection{Subspace variational quantum simulator}\label{sec:svqs}
\begin{figure}
	\centering
	\includegraphics[width=0.4\textwidth]{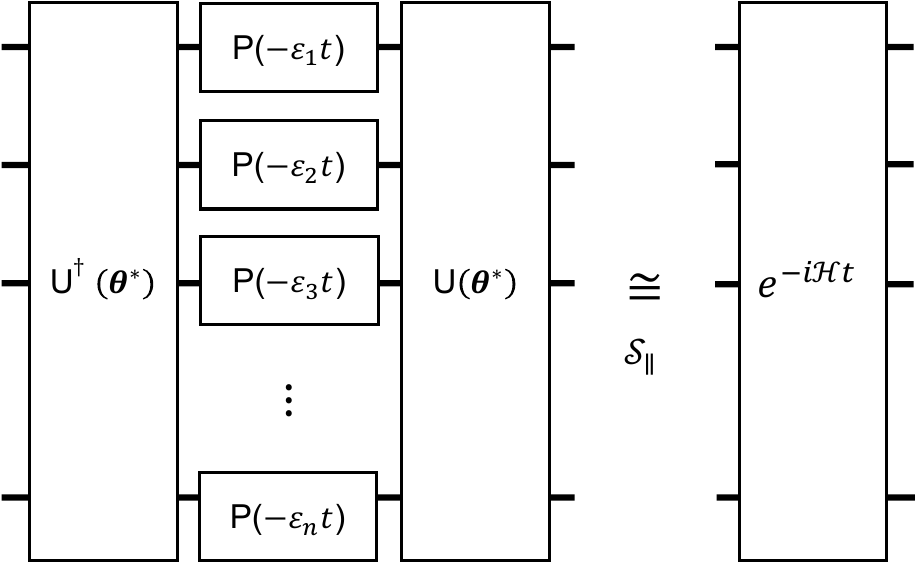}
	\caption{Quantum circuit for SVQS.
	$P(\xi)$ denotes a phase gate with the rotation angle $\xi$.
	The quantum circuit shown on the left approximates the time-evolution operator within the eigensubspace $\mathcal{S}_{\mathrm{\parallel}}$.}
	\label{circuit_svqs}
\end{figure}

The key idea of SVQS is to map a low-lying eigensubspace $\mathcal{S}_{\mathrm{\parallel}}$ of the target Hamiltonian $\mathcal{H}$ to a computational subspace $\mathcal{S}_{\mathrm{com}}$ spanned by the orthogonal initial states specified in SSVQE.
The procedure of the SVQS is summarized as follows:
\begin{enumerate}
	\item Construct a parameterized quantum circuit
	$U(\bm{\theta})$ and prepare $l$ input states $\{\ket{\varphi_j}|\ket{\varphi_j}=X_j\ket{0}^{\otimes n}\}_{j=1}^l\ (l\leq n)$ such that they are orthogonal to each other ($\braket{\varphi_j|\varphi_{j'}}=\delta_{j,j'}$).
	\item Minimize
	$\mathcal{L}_{\bm{\omega}}(\bm{\theta})=\sum_{j=1}^{l} \omega_{j}\braket{\varphi_{j}|U^{\dagger}(\bm{\theta})\mathcal{H}U(\bm{\theta})|\varphi_{j}}$, where the weight vector $\bm{\omega}$ is chosen such that $\omega_{i}>\omega_{j}$ when $i>j$. 
	\item After convergence, get the $j$th eigenenergy $E_j$ as $\bra{\varphi_j}U^\dagger(\bm{\theta}^*)\mathcal{H}U(\bm{\theta}^*)\ket{\varphi_j}$ with the converged variational parameter $\bm{\theta}^*=\argmin \mathcal{L}_{\bm{\omega}}(\bm{\theta})$.
	\item Prepare an initial state $\ket{\psi_{\mathrm{in}}}$ in the eigensubspace $\mathcal{S}_{\mathrm{\parallel}}$.
	\item Encode the input state $\ket{\psi_{\mathrm{in}}}$ on the eigensubspace $\mathcal{S}_{\mathrm{\parallel}}$ into the computational subspace $\mathcal{S}_{\mathrm{com}}$ by applying the Hermitian conjugate of the obtained circuit $U^\dagger(\bm{\theta}^*)$.
	\item Apply a single-qubit phase-rotation on each qubit, namely, $V(t)=\bigotimes_{j=1}^{l} P (-E_j t)$, where $P(\xi)=\ket{0}\bra{0}+e^{i\xi}\ket{1}\bra{1}$ is a phase gate.
	\item Decode the state $V(t)U^\dagger (\bm{\theta}^*)\ket{\psi_{\mathrm{in}}}$ on the computational subspace $\mathcal{S}_{\mathrm{com}}$ into the eigensubspace $\mathcal{S}_{\mathrm{\parallel}}$ by applying $U(\bm{\theta}^*)$. See also \cref{circuit_svqs} showing the quantum circuit corresponding to steps~4--6 of the procedure.
\end{enumerate}

Let us explain how the above procedure works to simulate the time evolution of a quantum system.
The circuit $U(\bm{\theta}^*)$ obtained in step~2 corresponds to a map between $\mathcal{S}_{\mathrm{\parallel}}$ and $\mathcal{S}_{\mathrm{com}}$.
Therefore, the circuit $U(\bm{\theta}^*)$ can be denoted as follows:
\begin{align}\label{eq_u}
U(\bm{\theta}^*)=\sum_j e^{i\delta_j}\ket{E_j}\bra{\varphi_j}+U_{\perp},
\end{align}
where $\{\delta_j\}_{j=1}^{l}$ are unknown phase offsets and $U_{\perp}$ is an unknown map between the subspaces complementary to $\mathcal{S}_{\mathrm{\parallel}}$ and $\mathcal{S}_{\mathrm{com}}$.
The unknown phase offset occurs due to the nature of SSVQE only evaluating the energy values.
This offset makes phase synchronization between the computational basis and eigenbasis difficult.
However, we can cancel this offset by utilizing the inversed parameterized quantum circuit, as described below.
The tensor product of the single-qubit phase rotations can be denoted as follows:
\begin{align}\label{eq_t}
V(t)
&=\sum_{j=1}^{l} e^{-iE_j t}\ket{\varphi_j}\bra{\varphi_j}+U'_{\perp}(t),
\end{align}
where $U'_{\perp}(t)$ is again a map between the complementary subspaces.
Under the conjugation by \cref{eq_u}, \cref{eq_t} transforms as follows:
\begin{align}\label{eq_time_evo}
\mathcal{T}(t)
&=U(\bm{\theta}^*)V(t)U^\dagger (\bm{\theta}^*)\\
&=\sum_{j=1}^{l} e^{-iE_j t}\ket{E_j}\bra{E_j} + U_{\perp}U'_{\perp}(t)U^\dagger_{\perp}.
\end{align}
\Cref{eq_time_evo} corresponds to the subspace time-evolution operator on the $l$-dimentional eigensubspace with an unknown time-dependent unitary operation on the complementary subspace.
The single-qubit phase rotations can be implemented with high-fidelity virtual $Z$ gates~\cite{mckay2017efficient}.
Note that when we know the expansion coefficients of the initial state on the eigenbasis, we can classically simulate the time evolution for the state by using the eigenenergies obtained in SSVQE.
However, it is generally hard to run state tomography for eigenstates with many qubits.
SVQS does not require state tomography of eigenstates, and can directly generate a time-evolution operator on a quantum circuit.


\subsection{Extensions of SVQS}
In this subsection, we propose extensions for SVQS.
The first extension is a method to extend the dimension of the simulatable eigensubspace $\mathcal{S}_{\mathrm{\parallel}}$.
In \cref{sec:svqs}, for a $n$-qubit target Hamiltonian, the dimension of the simulatable eigensubspace restricts to $n$, because there only exist $n$ one-hot states in the $n$-qubit system.
In this extension, we introduce $a$-ancilla qubits to extend the dimension of the simulatable eigensubspace to $n+a$.
The Hamiltonian of the $(n+a)$-qubit system is defined in terms of $n$-qubit Hamiltonian $\mathcal{H}_{n}$ as follows:
\begin{align}
\mathcal{H}=\mathcal{H}_{n} \otimes I_{a} - E_{\mathrm{B}} I_{n} \otimes \sum_{i=1}^{a} (I_i - Z_i),
\end{align}
where $I_{n,a}$ denote the identity operators for $n$ and $a$ qubit subspace, and $I_i$ and $Z_i$ denote the identity and Pauli-$Z$ operators on the $i$th ancilla qubit, and $E_{\mathrm{B}}$ is an energy constant much larger than the $(n+a)$th excited-state eigenenergy.
The first term of the target Hamiltonian corresponds to the problem Hamiltonian on the $n$-qubit system, and the second term corresponds to the static energy bias $E_{\mathrm{B}}$ applied along the $Z$-axis of the ancilla qubits.
If the energy bias $E_{\mathrm{B}}$ is larger than the eigenenergy of the $(n+a)$th excited eigenstate of the problem Hamiltonian, the $i$th excited eigenstate of the total Hamiltonian becomes 
\begin{align}
\ket{\psi^{i}}=\ket{\psi^{i}_{n}}\otimes\ket{0}^{\otimes a},
\end{align}
where $\ket{\psi^{i}_{n}}$ denotes the $i$th excited eigenstate of the $n$-qubit Hamiltonian $\mathcal{H}_{n}$.
Note that such a method does not disturb the low-lying eigenstates or eigenenergies of the problem Hamiltonian.
Therefore, by running SVQS with $n$ data qubits and $a$ ancilla qubit, we can search up to the $(n+a)$th excited state of the $n$-qubit Hamiltonian.

The second extension is controlled-SVQS.
By replacing the phase gates in the quantum circuits drawn in \cref{circuit_svqs} with controlled phase gates, the subspace time evolution obtained with SVQS can easily be extended to the controlled subspace time evolution.
Simulation of the controlled time evolution is useful for calculating the generalized Green functions~\cite{bonch2015green, abrikosov2012methods}, out-of-time-order correlations~\cite{Hashimoto_2017}, and Loschmidt echo signals~\cite{Wisniacki_2012}.

\section{Experiment}\label{sec:experiment}
\subsection{System}
Here, we demonstrate SVQS using two superconducting qubits~(Q0 and Q1) capacitively coupled with each other.
The qubits are a part of a $16$-qubit device~\cite{tamate2021scalable}.
\begin{table}
\caption{Parameters of the two superconducting transmon qubits~(Q0 and Q1) used in the experiments: the qubit frequency $\omega_\mathrm{q}$,  anharmonicity $\alpha$, relaxation time $T_1$, and echo dephasing time $T_2^\mathrm{echo}$.}
\begin{ruledtabular}
\begin{tabular}{ccccc}
    & $\omega_\mathrm{q}/2\pi$   & $\alpha/2\pi$         & $T_1$                 & $T_2^\mathrm{echo}$ \\ \hline
Q0  & $8.209~{\rm GHz}$ & $-380~{\rm MHz}$    & $4.3~{\rm \mu s}$   & $8.9~{\rm \mu s}$  \\
Q1  & $8.965~{\rm GHz}$ & $-410~{\rm MHz}$    & $4.6~{\rm \mu s}$   & $6.5~{\rm \mu s}$ \\
\end{tabular}
\end{ruledtabular}
\label{Tab:Experiment_parameter_field}
\end{table}
The parameters of the qubits are summarized in \Cref{Tab:Experiment_parameter_field}.
The static ZZ interaction strength between the qubits is $-230~\mathrm{kHz}$.

\begin{figure}
    \centering
    \includegraphics[width=0.4\textwidth]{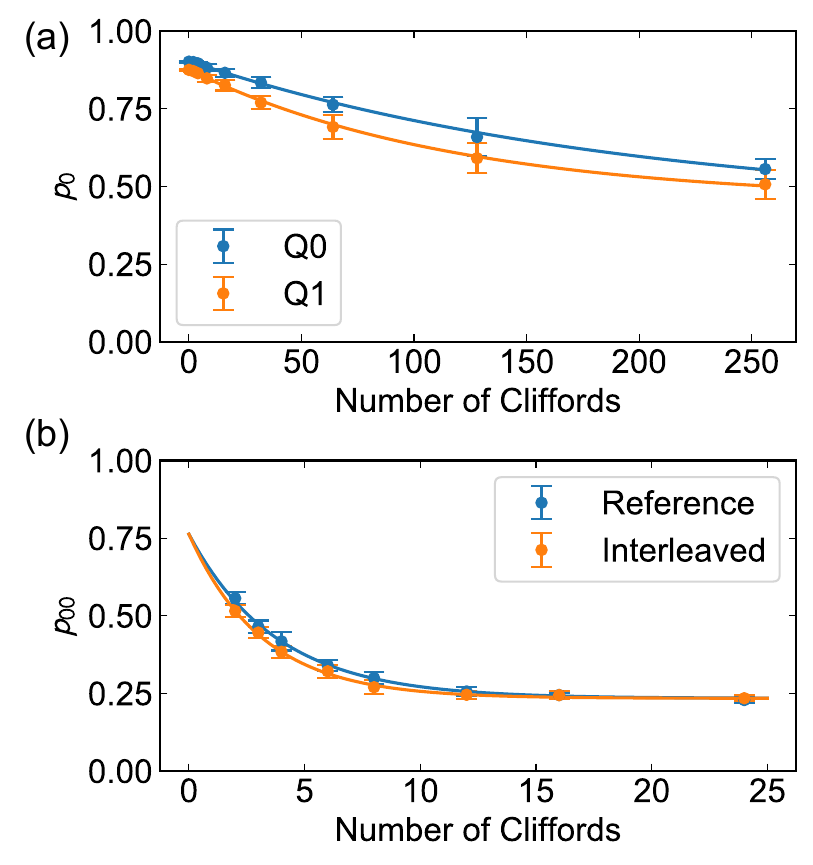}
    \caption{
    Experimental results of (a) the single-qubit simultaneous randomized benchmarking and (b) the two-qubit interleaved randomized benchmarking.
    To align the execution times of all the Clifford gates, the single-qubit and two-qubit Clifford gates are constructed based on Euler decomposition~\cite{mckay2017efficient} and KAK decomposition~\cite{e15061963}, respectively.
    Therefore, each single-qubit Clifford gate is implemented with two $R_{X}(\pi/2)$ and three $R_{Z}(\theta)$ gates, and each two-qubit Clifford gate is implemented with 28 $R_{X}(\pi/2)$, 27 $R_{Z}(\theta)$, and 6 $R_{ZX}(\pi/4)$ gates except the ones for the interleaved $R_{ZX}(\pi/2)$ gates.
    We take $10$ random circuits for each Clifford sequence length and have $10^4$ sampling of measurements for each random circuit to obtain a single data point.
    The ground state populations of each qubit $p_0$ and two qubits $p_{00}$ are fitted to the exponential decay.
    }
    \label{Fig:RB}
\end{figure}

We calibrate and implement single-qubit gates, $R_{Z}(\theta)=e^{-i\frac{\theta}{2}\mathrm{Z}}$ and  $R_{X}(\pi/2)=e^{-i\frac{\pi}{4}\mathrm{X}}$, and a two-qubit gate $R_{ZX}(\pi/4)=e^{-i\frac{\pi}{8}\mathrm{ZX}}$.
The $R_{Z}(\theta)$ gates are implemented with the virtual $Z$ gates~\cite{mckay2017efficient}.
The $R_{X}(\pi/2)$ gates are implemented with the shaped microwave pulse~\cite{PhysRevA.82.042339} with the total duration of $13.9~\mathrm{ns}$ for Q0 and $14.1~\mathrm{ns}$ for Q1.
The averaged gate fidelities of the single-qubit Clifford gates are $0.9974(4)$ and $0.9958(5)$ via the simultaneous randomized benchmarking~\cite{PhysRevLett.109.240504}, where the coherence limits of both qubits are $0.998$.
The $R_{ZX}(\pi/4)$ gates are implemented with a cross-resonance gate~\cite{rigetti2010fully, chow2011simple} with the total duration of $73.3~\mathrm{ns}$.
The averaged gate fidelity of the $R_{ZX}(\pi/2)$ gate is determined to be $0.963(4)$ via the interleaved randomized benchmarking~\cite{PhysRevLett.109.080505}, where the coherence limit is $0.961$.
The experimental results of the randomized benchmarking are shown in~\cref{Fig:RB}.

The multiplexed single-shot dispersive readout of the qubits~\cite{PhysRevA.69.062320, PhysRevLett.95.060501} are performed via readout resonators at $10.119~{\rm GHz}$ and $10.341~{\rm GHz}$ for Q0 and Q1, respectively, and a common Purcell filter~\cite{PhysRevLett.112.190504, PhysRevA.92.012325}.
The readout signal is amplified with an impedance-matched Josephson parametric amplifier~\cite{doi:10.1063/1.4886408, urade2020impedance}.
The thermal populations of the qubits in equilibrium are $7.84~\%$ and $11.72~\%$.
The assignment fidelities of the single-shot readout are $0.946$ and $0.922$, respectively.

\subsection{Subspace-search variational quantum eigensolver}
First, we demonstrate SSVQE.
The Hamiltonian of a hydrogen molecule in STO-3G basis can be converted into a two-qubit Hamiltonian~\cite{o2016scalable} as follows:
\begin{align}
\mathcal{H} = c_0II + c_1 ZI + c_2 IZ + c_3 XX + c_4 YY + c_5 ZZ,
\end{align}
where the coefficients $c_i$ are calculated with openfermion~\cite{mcclean2017openfermion} and Psi4~\cite{Parrish2017}.
In the following experiments, we use a minimal clique cover~\cite{doi:10.1021/acs.jctc.0c00008,jena2019pauli,doi:10.1063/1.5141458,gokhale2019minimizing,crawford2021efficient} to reduce the number of experiments required for the evaluation of the expectation values of the Hamiltonian.

\begin{figure}
    \centering
    \includegraphics[width=0.45\textwidth]{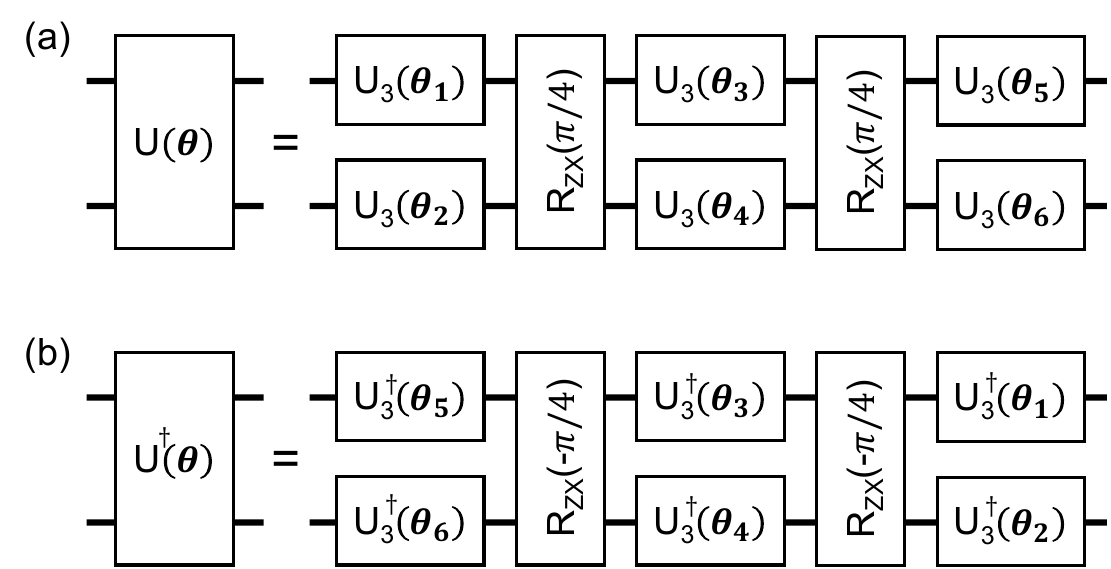}
    \caption{
    (a)~Hardware efficient ansatz used in the SSVQE experiments and (b)~its Hermitian conjugate used in the SVQS experiments, where $U_3(\bm{\theta})$ ($\bm{\theta} = \{ \bm{\theta_i} \}$; $ i =1, \dots, 6$) represents a parameterized single-qubit rotation implemented with two $R_{X}(\pi/2)$ gates and three parameterized $RZ$ gates~\cite{mckay2017efficient}.
    Each $\bm{\theta_i}$ consists of three phase parameters.
    }
    \label{Fig:ansatz}
\end{figure}

In the optimization protocol, we employ the hardware-efficient ansatz $U(\bm{\theta})$ with $18$ parameters shown in~\cref{Fig:ansatz}(a).
We optimize the parameterized quantum circuit on our experimental system to incorporate the system imperfection with the initial parameters given by the numerical simulation.
We prepare the initial states by thermalizing the qubits to the equilibrium for a sufficiently longer time than the energy relaxation time and then performing an X-gate on one of the qubits.
Then, we operate the parameterized quantum circuit and evaluate the energy expectation values for the final states.
Finally, we calculate the cost function by taking the linear sum of the energy expectation values for the final states with the weighting factors $\omega_0=2$ and $\omega_1=1$.

To minimize the effect of imperfections in the circuit, we use gate-error mitigation~\cite{PhysRevX.7.021050, temme2017error, Endo_2021} and measurement-error mitigation~\cite{Maciejewski_2020, PhysRevA.100.052315, Kwon_2021, Endo_2021}.
The former is a method to estimate the measurement results for the ideal case without gate errors by the linear extrapolation of experimental results obtained for different gate execution times.
The latter is a method to estimate the ideal measurement results in the absence of measurement errors by using a predetermined measurement confusion matrix.
We run the experimental SSVQE with $36$~iterations.
Both in the numerical and experimental SSVQEs, we adopt the sequential minimal optimization method~\cite{knaka_opt} as the optimizer.

\begin{figure}
    \centering
    \includegraphics[width=0.4\textwidth]{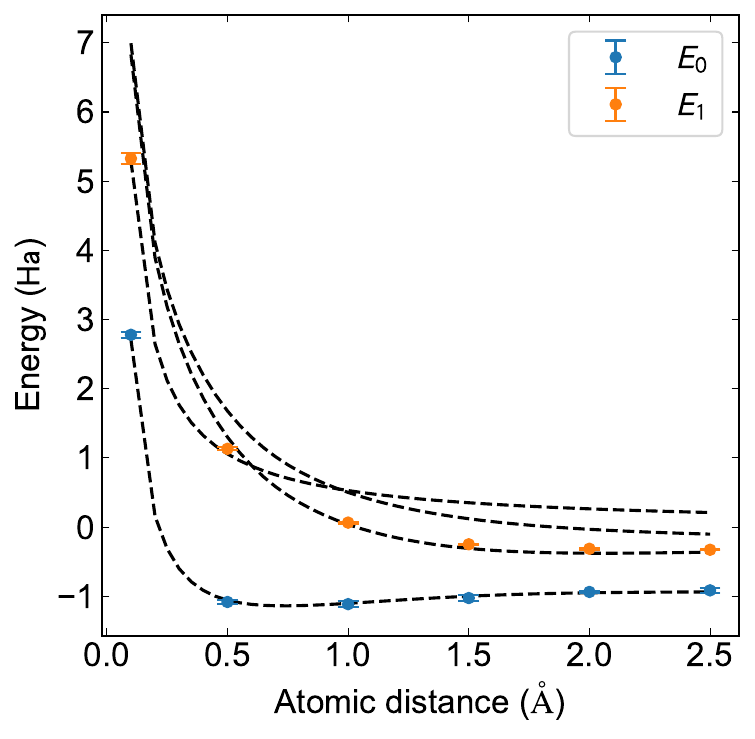}
    \caption{
    SSVQE for different atomic distances.
    The blue and orange dots represent the experimentally obtained ground and first-excited eigenenergies, respectively.
    The error bars indicate the standard deviation of the eigenenergies, where the shot number is $10^4$.
    The black dashed lines depict the theoretical eigenenergies of the hydrogen molecule.
    }
    \label{Fig:SSVQE}
\end{figure}
\Cref{Fig:SSVQE} shows the experimental results of the SSVQE for different atomic distances, where the evaluated eigenenergies match well to the theoretical values almost within the standard deviation range.
The details of the optimization traces are shown in \cref{opt_trace}.

\subsection{Subspace variational quantum simulator}
Next, we demonstrate SVQS.
In the following experiments, the atomic distance is set to $1.0~\mbox{\AA}$, for simplicity.

As discussed in \cref{sec:svqs}, the subspace time-evolution operator for the low-lying eigensubspace $\mathcal{S}_{\mathrm{\parallel}}$ can be systematically implemented with the quantum circuit drawn in \cref{circuit_svqs}, where $U(\bm{\theta}^*)$ obtained from SSVQE~[\cref{Fig:ansatz}(a)] and its Hermitian conjugate $U^\dagger (\bm{\theta}^*)$~[\cref{Fig:ansatz}(b)] are used.

We perform quantum process tomography~\cite{Mohseni_2008} for the subspace time-evolution operators mimicked by SVQS, where we apply the gate-error mitigation and the measurement-error mitigation.
To estimate the quantum processes, we use Quara~\cite{quara}, a software package for quantum characterization.
In Quara, quantum processes are estimated with the maximum likelihood method under a predetermined measurement confusion matrix.
We sweep the evolution time from $0$ to $5.4~h/{\rm Ha}$, which roughly corresponds to one cycle of the time evolution, $T=2\pi/(E_1 - E_0)$, in the low-lying eigensubspace.

To properly evaluate the performance of the time evolution mimicked by SVQS, we introduce the subspace process infidelity~(SPIF) as a measure of the difference between two quantum processes in a subspace.
SPIF is a natural derivative of process infidelity~(PIF) used for quantum processes in the whole space.
The details of the definition of SPIF are described in \Cref{sec:spf}.
\begin{figure}
    \centering
	\includegraphics[width=0.4\textwidth]{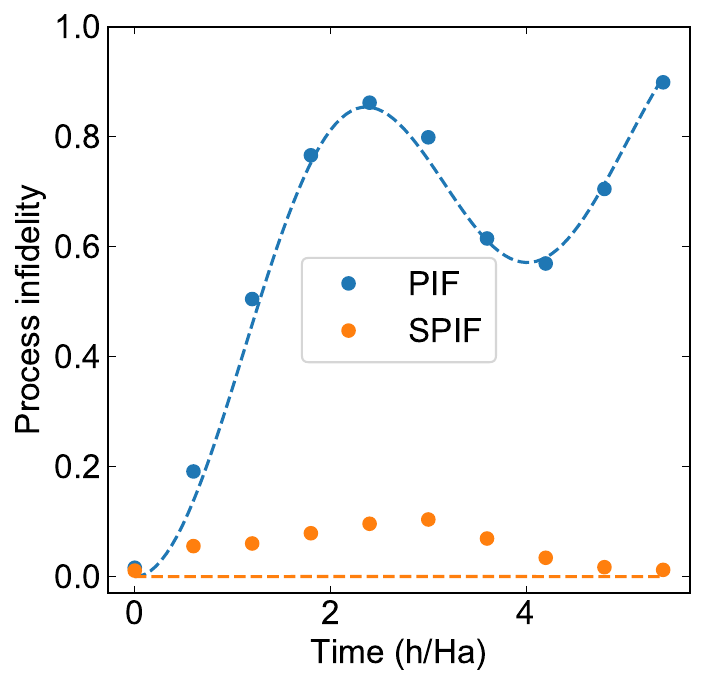}
    \caption{
    SVQS for the evolution time from $0$ to $5.4~h/{\rm Ha}$.
    The blue and orange dashed lines represent the PIF and SPIF between the ideal and numerically calculated time-evolution operators mimicked by SVQS, respectively.
    The data points represent the corresponding experimental results.
    }
    \label{Fig:SVQS}
\end{figure}
\Cref{Fig:SVQS} shows the experimental results of SVQS.
Since SVQS mimics the dynamics only within the low-lying eigensubspace, only the SPIF becomes zero, and the PIF varies with time theoretically.
The experimental SPIF and PIF show similar characteristics to the numerical ones.
The experimental SPIF takes values of $0.02$--$0.12$.
In the two-dimensional low-lying eigensubspace, subspace time evolution corresponds to a rotating operation in $SO(3)$.
A detailed analysis in \Cref{sec:sptm} reveals the subspace time evolution corresponding to a $1.1\%$ speed error and a $19^{\circ}$ axis error in the rotating operation.

\section{Conclusion}\label{sec:conclusion}
In this paper, we proposed an efficient algorithm, subspace variational quantum simulator~(SVQS), to simulate quantum dynamics on a NISQ device.
The circuit depth required in SVQS is at most twice as large as that of the subspace-search variational quantum eigensolver~(SSVQE) on the same device.
Recently, there have been proposals to implement VQE for larger-scale quantum systems~\cite{yoshioka2020variational, PhysRevLett.127.040501, fujii2020deep, cao2021larger, yamamoto2021quantum}.
It is expected that SVQS can also be implemented for larger-scale systems.

We experimentally demonstrated SVQS using a system consisting of two superconducting qubits.
Even with the limited device performance, we approximated the quantum dynamics in the two-dimensional low-lying eigensubspace of a hydrogen molecule with subspace process fidelity of $0.88$--$0.98$.
This suggests that our proposal is effective for experiments on NISQ devices because of the modest requirements for experimental devices.

In SVQS, we apply the parameterized quantum circuit obtained with SSVQE as a map between a low-lying eigensubspace and a computational subspace and implement subspace time evolution as phase rotations on the eigenbasis.
The relationship between SSVQE, which gives the desired map, and SVQS, which utilizes the map, is similar to the relationship between quantum phase estimation~\cite{PhysRevA.54.4564, cleve1998quantum, PhysRevLett.82.2207} and HHL algorithm~\cite{PhysRevLett.103.150502}.
The similarity suggests a possible path toward hybrid quantum--classical algorithms that can be applied to a wider range of problems including linear equations.

\section{Data availability}
The data that support the findings of this study are available from the corresponding authors upon reasonable request.

\section{Code availability}
The code that is deemed central to the conclusions are available from the corresponding author upon reasonable request.

\section{Acknowledgements}\label{sec:acknowledgements}
This work was supported by QunaSys.
K.H., K.M.N, and K.M.\ thanks IPA for its support through MITOU Target program.
K.H.\ is supported by JSPS KAKENHI (No.~JP21J15221).
K.M.N\ is supported by JSPS KAKENHI (No.~JP20J13955) and the Daikin Endowed Research Unit: “Research on Physics of Intelligence”, School of Science, the University of Tokyo.
K.M.\ is supported by JST PRESTO (No.~JPMJPR2019) and JSPS KAKENHI
(No.~JP20K22330).
K.F.\ is supported by JSPS KAKENHI (No.~JP16H02211), JST ERATO (No.~JPMJER1601), JST CREST (No.~JPMJCR1673), and MEXT Q-LEAP (No.~JPMXS0120319794).
Y.S.\ is supported by JST PRESTO (No.~JPMJPR1916) and JST Moonshot R\&D (No.~JPMJMS2061).
K.H., Y.S., Z.Y., K.Z., T.S., S.T., Y.T., and Y.N.\ are partly supported by JST ERATO (No.~JPMJER1601) and by MEXT Q-LEAP (No.~JPMXS0118068682).

\section{Author contributions}
K.H., K.M.N, and K.M designed the theoretial concepts.
K.H. designed and performed the experiments.
S.T. and Y.T designed the device and the control system.
Z.Y. and K.Z. fabricated and calibrated the impedance-matched Josephson parametric amplifier.
K.H. performed the numerical and analytical calculations.
K.H., Y.S., and T.S., analysed the data.
K.H. wrote the manuscript with feedback from the other authors.
K.F. and Y.N. supervised the project.

\section{Competing interests}
The authors declare no competing interests.

\bibliography{bibliography}

\appendix
\begin{widetext}

\section{Detailed analysis of experimental results in SSVQE}\label{opt_trace}
In SSVQE, we use the gate-error and measurement-error mitigations.
To mitigate gate errors, we scale the effective gate time by a factor of $1$, $1.5$, and $2$, respectively, by adding a necessary delay time to both ends of a control microwave pulse for $R_{X}(\pi/2)$ and $R_{ZX}(\pi/4)$.
We conduct experiments for the three cases. 
Then, we linearly extrapolate the experimental results to the scale factor of $0$ and obtained the mitigated results.
\begin{figure}
    \centering
	\includegraphics[width=0.9\textwidth]{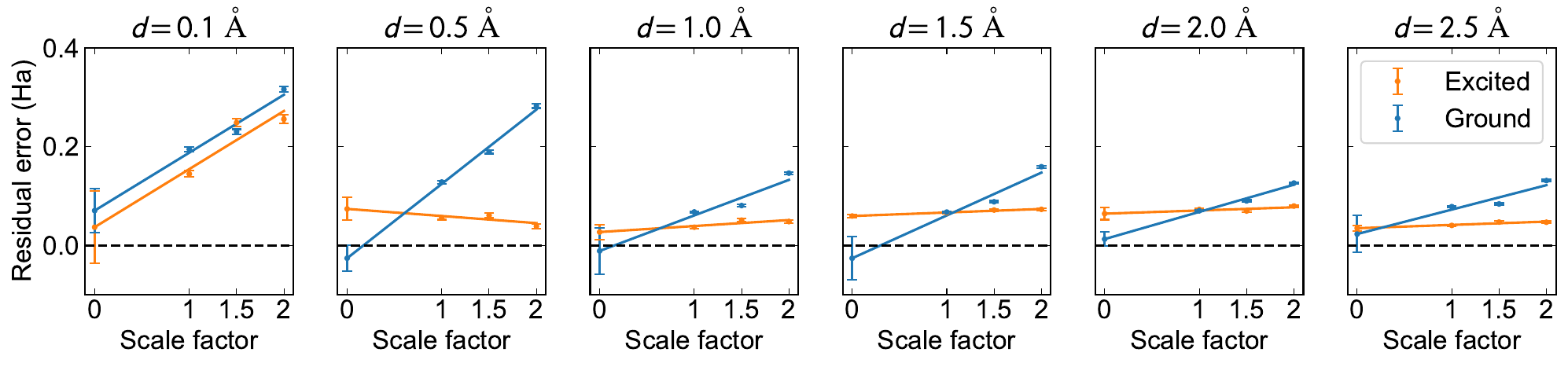}
	\caption{
	Gate-error mitigation results of SSVQE after convergence for different atomic distances $d$.
	The blue and orange dots represent the residual errors of the ground and first-excited energies, respectively.
	The horizontal axis represents the scale factor corresponding to the additional delay time for gate-error mitigation.
	The error bars indicate the standard deviation of the eigenenergies, where the shot number is $10^4$.
	}
    \label{Fig:ssvqe_lmit}
\end{figure}
\Cref{Fig:ssvqe_lmit} shows the experimental results with the gate-error mitigation at the convergence point in SSVQE.
The linear extrapolation for the evaluated ground-state energies works well, and the mitigated ground-state energies match theoretical values within the standard deviation.
In contrast, except for the case with the atomic distance of $0.1~\mbox{\AA}$, the evaluated first-excited-state energies are not much affected by the insertion of the delay time.
The mitigated first-excited-state energies also deviate slightly from the theoretical values.
We guess that the deviation of the first-excited-state energies from the theoretical values originates in a bottleneck for the convergence other than the incoherent error during the pulse sequences.
As shown in \cref{Fig:SSVQE}, while the ground-state energies are well isolated from the other eigenenergies, the first-excited-state energies are close to the second-excited-state energies.
Only in the case with the atomic distance of $0.1~\mbox{\AA}$, the first-excited-state energy is sufficiently isolated from the second-excited-state energy, and thus the gate-error mitigation works well.
We conclude that the proximity of the first and the second-excited-state energies slows the convergence of SSVQE, resulting in a slight contamination of the second excited state in the final states.

To mitigate the measurement error, we use a predetermined measurement confusion matrix $C$.
We estimate the true histogram $x$ which minimizes $|y-Cx|^2$ for the measurement histogram $y$ obtained in the experiments.
Here, we used SLSQP~\cite{scipy} as an optimizer for the minimization.

\begin{figure}
    \centering
	\includegraphics[width=0.9\textwidth]{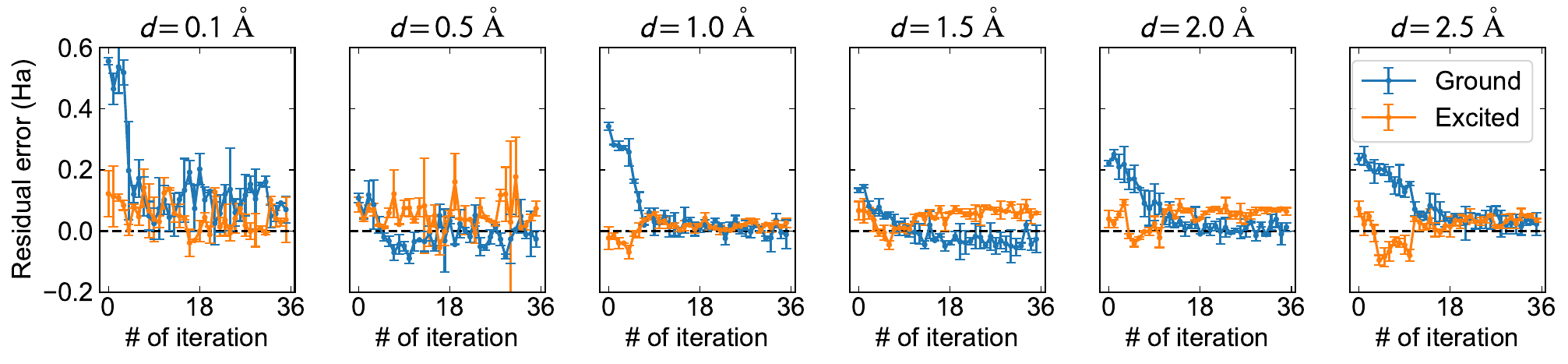}
	\caption{
	Optimization traces in SSVQE for different atomic distances $d$.
	The blue and orange dots represent the residual errors of the ground- and first-excited-state energies, respectively.
	The error bars indicate the standard deviation of the eigenenergies, where the shot number is $10^4$.
	}
    \label{Fig:trace_ssvqe}
\end{figure}
\Cref{Fig:trace_ssvqe} shows the experimental traces of the residual errors in SSVQE converging to the low-lying eigenenergies for different atomic distances, where the residual errors almost converge after $36$ iterations.
In the optimization traces, some of the evaluated ground-state energies are less than the true ground-state energies, which is presumably due to the linear extrapolation in the gate-error mitigation.
With the averaging of $10^4$ times, the shot-noise of each Pauli observable is at most $(1/\sqrt{2})\times10^{-2}\sim0.007$.
As shown in~\cref{Fig:ssvqe_lmit}, the larger error bars mainly arise from the fitting in the linear extrapolation.
We also find that the error bar for $d\leq0.5~\mbox{\AA}$ is larger on average than that for $d\geq1.0~\mbox{\AA}$.
As mentioned above, the larger error here is mainly due to the fact that the experimental results obtained for the gate-error mitigation no longer follow the linear model, suggesting that the nature of the error in the ansatz circuit has changed. 

As shown in \cref{Fig:SSVQE}, there is a structural transition at $d\sim0.6~\mbox{\AA}$, where the first and second excited energies intersect with each other.
Here, we summarize coefficients of the Hamiltonian for hydrogen molecule at each atomic distance in \cref{Tab:hamiltonian_parameter_field}.
\begin{table}
\caption{List of coefficients $\{c_i\}_{i=0}^5$ of the Hamiltonian for hydrogen molecule at each atomic distance $d$.}
\begin{ruledtabular}
\begin{tabular}{ccccccc}
$d$ & $c_0$ & $c_1$ & $c_2$ & $c_3$ & $c_4$ & $c_5$ \\ \hline
$0.1~\mathrm{\AA}$ & $5.46~\mathrm{Ha}$  & $0.6~\mathrm{Ha}$   & $-1.45~\mathrm{Ha}$ & $0.69~\mathrm{Ha}$  & $0.08~\mathrm{Ha}$  & $0.08~\mathrm{Ha}$  \\
$0.5~\mathrm{\AA}$ & $0.75~\mathrm{Ha}$  & $0.43~\mathrm{Ha}$  & $-0.74~\mathrm{Ha}$ & $0.62~\mathrm{Ha}$  & $0.08~\mathrm{Ha}$  & $0.08~\mathrm{Ha}$  \\
$1.0~\mathrm{\AA}$ & $-0.01~\mathrm{Ha}$ & $0.27~\mathrm{Ha}$  & $-0.26~\mathrm{Ha}$ & $0.52~\mathrm{Ha}$  & $0.10~\mathrm{Ha}$  & $0.10~\mathrm{Ha}$  \\
$1.5~\mathrm{\AA}$ & $-0.21~\mathrm{Ha}$ & $0.19~\mathrm{Ha}$  & $-0.07~\mathrm{Ha}$ & $0.44~\mathrm{Ha}$  & $0.11~\mathrm{Ha}$  & $0.11~\mathrm{Ha}$  \\
$2.0~\mathrm{\AA}$ & $-0.27~\mathrm{Ha}$ & $0.13~\mathrm{Ha}$  & $0.01~\mathrm{Ha}$  & $0.39~\mathrm{Ha}$  & $0.13~\mathrm{Ha}$  & $0.13~\mathrm{Ha}$  \\
$2.5~\mathrm{\AA}$ & $-0.3~\mathrm{Ha}$  & $0.11~\mathrm{Ha}$  & $0.05~\mathrm{Ha}$  & $0.35~\mathrm{Ha}$  & $0.14~\mathrm{Ha}$  & $0.14~\mathrm{Ha}$
\end{tabular}
\end{ruledtabular}
\label{Tab:hamiltonian_parameter_field}
\end{table}
Before and after the transition, low-lying eigenstates differ from each other as follows:
\begin{align}
\ket{E_0}_{d\le 0.5~\mathrm{\AA}}\sim \ket{01},&\ \ket{E_1}_{d\le 0.5~\mathrm{\AA}}\sim \ket{00}, \\
\ket{E_0}_{d\ge 1.0~\mathrm{\AA}}\sim \ket{01},&\ \ket{E_1}_{d\ge 1.0~\mathrm{\AA}}\sim \ket{10}.
\end{align}
Therefore, the maps to be implemented with SSVQE also differ as follows:
\begin{align}
U_{d\le 0.5~\mathrm{\AA}} &\sim \ket{10}\bra{10} + e^{i\delta_1}\ket{00}\bra{01} + U_1^\perp \\
U_{d\ge 1.0~\mathrm{\AA}} &\sim \ket{10}\bra{10} + e^{i\delta_2}\ket{01}\bra{01} + U_2^\perp,
\end{align}
where $\delta_{1,2}$ and $U_{1,2}^\perp$ are unknown phase factors and unknown maps between the complementary subspaces, respectively.
While the latter map is an identity operator, the former map is an entangling operation that rotates a qubit according to the state of the other qubit.
Therefore, the two $R_{ZX}(\pi/4)$ gates contained in the ansatz circuit need to interfere with each other generatively in the former case and destructively in the latter case.
The noise of the ansatz circuit is mainly distributed to the two $R_{ZX}(\pi/4)$ gates, which account for $147~\mathrm{ns}$ of the total duration of the ansatz circuit of $213~\mathrm{ns}$.
Therefore, the change in the constraints imposed on the interference of the two $R_{ZX}(\pi/4)$ gates can significantly change the nature of the circuit noise.

\section{Subspace process fidelity and subspace Pauli transfer matrix}\label{sec:spf}
Process fidelity is a typical measure of the closeness between two quantum processes.
The definition of the process fidelity is given as follows:
\begin{align}
\mathcal{F}(\mathcal{U}, \mathcal{V})\equiv\frac{\mathrm{Tr}\left[\mathcal{S}^\dagger_\mathcal{U}\mathcal{S}_\mathcal{V}\right]}{d^2},
\end{align}
where $d$ is the dimension of the quantum processes, and $S_{\mathcal{U}, \mathcal{V}}$ is the superoperator representation on the maps $\mathcal{U}, \mathcal{V}$, respectively.
In this subsection, we introduce a new fidelity measure for quantum processes in subspaces, called subspace process fidelity.
The subspace process fidelity is defined as follows:
\begin{align}
\mathcal{F}_{\mathcal{D}_{\mathrm{i}}, \mathcal{D}_{\mathrm{o}}}(\mathcal{U}, \mathcal{V})
\equiv
\mathcal{F}(
\mathcal{D}_{\mathrm{o}} \circ \mathcal{U}\circ \mathcal{D}_{\mathrm{i}},
\mathcal{D}_{\mathrm{o}} \circ \mathcal{V}\circ \mathcal{D}_{\mathrm{i}}
),
\end{align}
where $\mathcal{D}_{i,o}$ are the relaxation operators for the input and output subspaces, respectively.
To evaluate the subspace time evolution in the low-lying eigensubspace, we introduce the subspace perfect depolarizing channel $\mathcal{D}_\mathrm{\perp}$ as follows:
\begin{align}
\mathcal{D}_\mathrm{\perp}(\rho)=\mathcal{P}_{\mathrm{\parallel}}(\rho) + \frac{\mathrm{Tr}\left[\mathcal{P}_{\mathrm{\perp}}(\rho)\right]}{d_{\mathrm{\perp}}}I_{\mathrm{\perp}},
\end{align}
where $\mathcal{P}_{\mathrm{\parallel, \perp}}$ are the projection operators on the low-lying eigensubspace and its complementary subspace, respectively, with the dimension $d_{\mathrm{\perp}}$ and the identity operator $I_{\mathrm{\perp}}$ on the complementary subspace.
The subspace process fidelity in the low-lying eigensubspace is calculated as $\mathcal{F}_{\mathcal{D}_{\mathrm{\perp}}, \mathcal{D}_{\mathrm{\perp}}}(\mathcal{U}, \mathcal{V})$.
The subspace process infidelity in the low-lying eigensubspace is also calculated as $1-\mathcal{F}_{\mathcal{D}_{\mathrm{\perp}}, \mathcal{D}_{\mathrm{\perp}}}(\mathcal{U}, \mathcal{V})$.
In addition, we introduce the subspace Pauli transfer matrix representation to visualize the properties of the quantum processes in the subspace defined as follows:
\begin{align}
\mathcal{R}^{\mathrm{\parallel}}_{ij}
=
\frac{
\mathrm{Tr}
\left[
\sigma_i^{\mathrm{\parallel}}
\mathcal{T}(t)
\sigma_j^{\mathrm{\parallel}}
\right]}
{d_{\mathrm{\parallel}}},
\label{Eq:sptm}
\end{align}
where $\mathcal{R}^{\mathrm{\parallel}}_{ij}$ is the $(i,j)$ element of the subspace Pauli transfer matrix, and $\sigma_i^{\mathrm{\parallel}}$ is the $i$th subspace Pauli operator.
For the two-dimensional low-lying eigensubspace of a hydrogen molecule, the subspace Pauli operators are defined as follows:
\begin{align}
\sigma_0^{\mathrm{\parallel}} &= \ket{E_0}\bra{E_0} + \ket{E_1}\bra{E_1} \\
\sigma_1^{\mathrm{\parallel}} &= \ket{E_1}\bra{E_0} + \ket{E_0}\bra{E_1} \\
\sigma_2^{\mathrm{\parallel}} &= i(\ket{E_1}\bra{E_0} - \ket{E_0}\bra{E_1}) \\
\sigma_3^{\mathrm{\parallel}} &= \ket{E_0}\bra{E_0} - \ket{E_1}\bra{E_1}.
\end{align}
and $d_{\mathrm{\parallel}}=2$ is the dimension of the low-lying eigensubspace.
Note that the interaction between the target subspace and the complementary subspace is regarded as the leakage, and the obtained subspace Pauli transfer matrix generally does not preserve the trace of the system.
In the low-lying eigensubspace, the Hamiltonian of a hydrogen molecule is given as follows:
\begin{align}
\mathcal{H}^{\mathrm{\parallel}}=
E_0\ket{E_0}\bra{E_0}+E_1\ket{E_1}\bra{E_1}
=
\frac{E_0+E_1}{2}\sigma_0^{\mathrm{\parallel}}
+
\frac{E_0-E_1}{2}\sigma_3^{\mathrm{\parallel}}.
\end{align}
Therefore, the subspace time-evolution operator corresponds to a phase rotation gate in the low-lying eigensubspace as follows:
\begin{align}
\mathcal{T}^{\mathrm{\parallel}}(t)=\exp{\left\{-i\left(\frac{E_0-E_1}{2}t\right) \sigma_3^{\mathrm{\parallel}}\right\}}.
\end{align}

\section{Detailed analysis of experimental results in SVQS}\label{sec:sptm}
The experimentally obtained Pauli transfer matrices of the subspace time-evolution operator mimicked by SVQS are shown in \cref{Fig:ptm_exp}.
\begin{figure}
    \centering
	\includegraphics[width=0.9\textwidth]{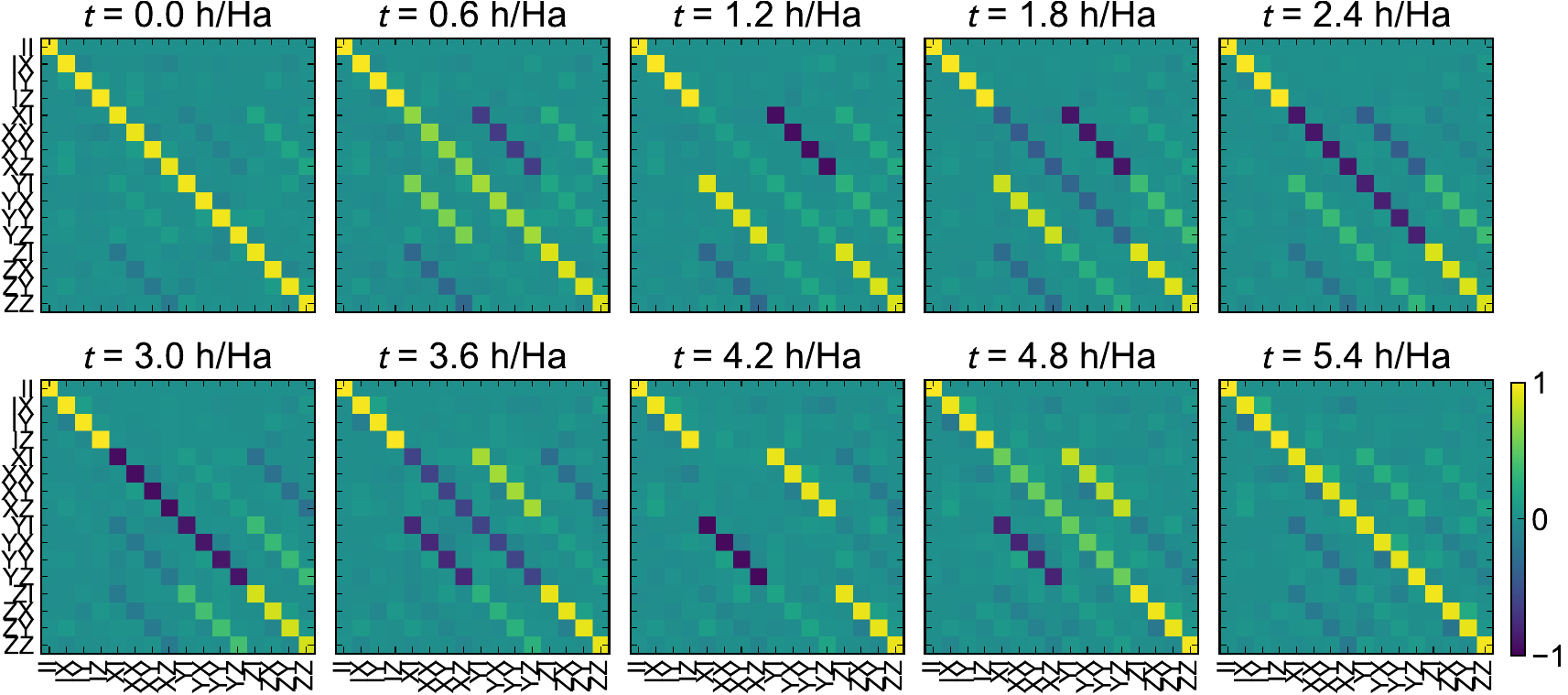}
	\caption{
	Experimentally obtained Pauli transfer matrices of time-evolution operator mimicked by SVQS for different evolution times $t$.
	}
    \label{Fig:ptm_exp}
\end{figure}
The Pauli transfer matrices are estimated from the gate-error-mitigated results with the maximum likelihood estimation using the predetermined initial thermal populations and measurement confusion matrix, which is supported by a quantum characterization toolkit, Quara~\cite{quara}.
\Cref{Fig:ptm_num} shows the theoretical Pauli transfer matrices of the ideal time-evolution operator.
\begin{figure}
    \centering
    \includegraphics[width=0.9\textwidth]{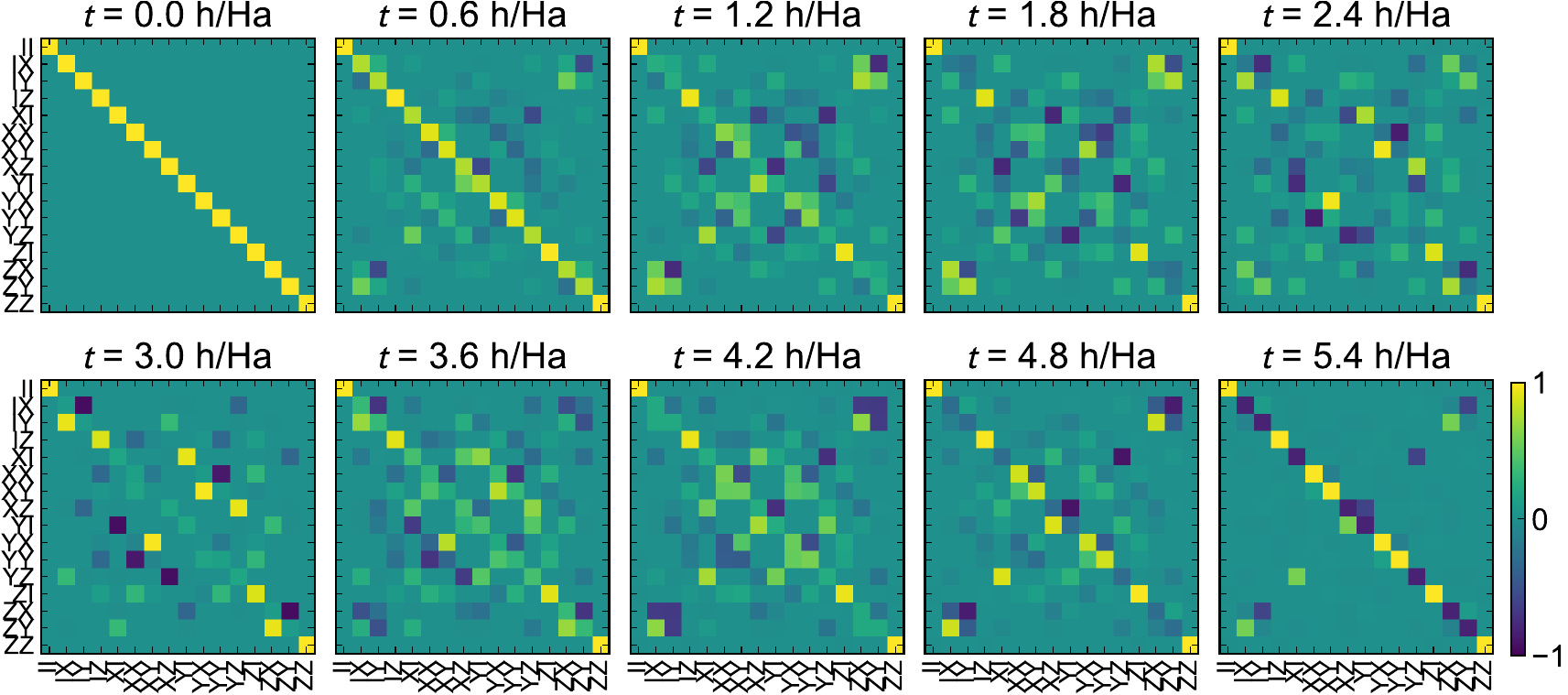}
    \caption{
    Numerically obtained Pauli transfer matrices of the time-evolution operator of a hydrogen molecule for different evolution times $t$.
    }
    \label{Fig:ptm_num}
\end{figure}
At first glance, the Pauli transfer matrices shown in Figs.~\ref{Fig:ptm_exp} and \ref{Fig:ptm_num} look totally different.
This is because SVQS mimics only the subspace time-evolution operator within the low-lying eigensubspace $\mathcal{S}_{\mathrm{\parallel}}$.
Therefore, for a proper comparison, we extract the subspace Pauli transfer matrices of the time-evolution operators in the low-lying eigensubspace defined in \cref{Eq:sptm}.
\begin{figure}
    \centering
    \includegraphics[width=0.9\textwidth]{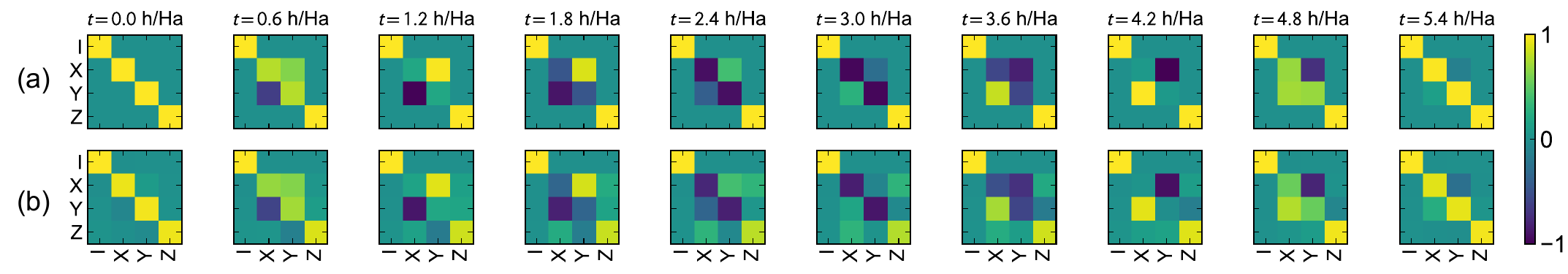}
	\caption{
	(a)~Numerically and (b)~experimentally obtained subspace Pauli transfer matrices of the subspace time-evolution operator in the low-lying eigensubspace $\mathcal{S}_\mathrm{\parallel}$ for the evolution time $t=0$ to $5.4~h/{\rm Ha}$.
	}
    \label{Fig:sptm}
\end{figure}
\Cref{Fig:sptm} shows the numerically and experimentally obtained subspace Pauli transfer matrices of the time-evolution operator in the low-lying eigensubspace $\mathcal{S}_\mathrm{\parallel}$ for different evolution times $t$.
From the subspace Pauli transfer matrices, we can estimate the Hamiltonian mimicked by SVQS.
Here, we use the dynamics generator analysis~\cite{sugiyama2020reliable} to extract unitary components from the quantum processes. 
For the extracted unitary components, we numerically search for the approximate Hamiltonian of subspace time-evolution operator mimicked by SVQS with Powell method~\cite{powell1964efficient}.
The ideal and fitted Hamiltonian in the low-lying eigensubspace is written as follows:
\begin{align}
\mathcal{H}_{\mathrm{ideal}}^{\mathrm{\parallel}} &\sim+0.00\sigma_1^{\mathrm{\parallel}} + 0.00\sigma_2^{\mathrm{\parallel}}-0.57\sigma_3^{\mathrm{\parallel}}\\
\mathcal{H}_{\mathrm{fit}}^{\mathrm{\parallel}} &\sim-0.19\sigma_1^{\mathrm{\parallel}} + 0.01\sigma_2^{\mathrm{\parallel}}-0.54\sigma_3^{\mathrm{\parallel}}.
\end{align}
The fitted Hamiltonian approximates the unitary components of the subspace time evolution with a process fidelity of $0.998(1)$.
From the fitted Hamiltonian, we find that the time evolution mimicked by SVQS has a rotation-speed error of $1.1\%$ and a rotation-axis error of $19.3^{\circ}$.

\end{widetext}
\end{document}